\documentclass[conference]{IEEEtran}
\usepackage{cite}
\usepackage{xspace}
\usepackage{amsmath,amssymb,amsfonts}
\usepackage{algorithmic}
\usepackage{graphicx}
\usepackage{textcomp}
\usepackage{xcolor}
\usepackage{fancyhdr}
\usepackage[hyphens]{url}

\usepackage{amsmath,amssymb,amsfonts}  %includes for \ding command
\usepackage{pifont} %includes for \ding command

\usepackage{hyperref}

\newcommand{\ourframework}{Accel-GCN\xspace} %\xspace

\def\BibTeX{{\rm B\kern-.05em{\sc i\kern-.025em b}\kern-.08em
    T\kern-.1667em\lower.7ex\hbox{E}\kern-.125emX}}
\usepackage{booktabs}
\usepackage{array}
\usepackage{amsfonts,amssymb}
\usepackage{graphicx}
\usepackage{amsmath} % for \textcircled{}
\usepackage{xcolor} % for \color and \textcolor
\usepackage{algorithmic}
\usepackage{algorithm}
\algsetup{indent=0.5em}
\newcommand{\minuseq}{\mathrel{-}=}

\usepackage{listings}

\usepackage[symbol]{footmisc}

\usepackage{tikz}
\usepackage{cite}
\usepackage{multirow}
\algsetup{linenosize=\footnotesize}

\begin{document}

%GNNAdvisor: An Adaptive and Efficient Runtime System for GNN Acceleration on GPUs
%Sextans: A Streaming Accelerator for General-Purpose Sparse-Matrix Dense-Matrix Multiplication
%GraphLily: Accelerating Graph Linear Algebra on HBM-Equipped FPGAs  
%High-Performance Sparse Linear Algebra on HBM-Equipped FPGAs Using HLS: A Case Study on SpMV
\title{\ourframework: High-Performance GPU Accelerator Design for Graph Convolution Networks\\
}
% \title{A new way of GPU acceleration for Graph Neural Networks\\
% % {\footnotesize \textsuperscript{*}Note: Sub-titles are not captured in Xplore and
% % should not be used}
% % \thanks{Identify applicable funding agency here. If none, delete this.}
% }

% \and

\author{\IEEEauthorblockN{ \textsuperscript{$\star$}Xi Xie\textsuperscript{[1]}, \textsuperscript{$\star$}Hongwu Peng\textsuperscript{[1]}, Amit Hasan\textsuperscript{[1]}, Shaoyi Huang\textsuperscript{[1]}, Jiahui Zhao\textsuperscript{[1]}, \textsuperscript{$\dagger$}Haowen Fang, Wei Zhang\textsuperscript{[1]}, \\ Tong Geng\textsuperscript{[2]}, Omer Khan\textsuperscript{[1]}, and Caiwen Ding\textsuperscript{[1]}}
\IEEEauthorblockA{
\textsuperscript{$\star$}These authors contributed equally. \\
\textsuperscript{[1]}University of Connecticut, 
\textsuperscript{[2]}University of Rochester \\
\textsuperscript{[1]}\{xi.xie, hongwu.peng, amit.hasan, shaoyi.huang, jiahui.zhao, wei.13.zhang, khan, caiwen.ding\}@uconn.edu, \\ 
\textsuperscript{[2]} haowfang@gmail.com, 
\textsuperscript{[3]}tgeng@ur.rochester.edu
}
}
\maketitle

%\textsuperscript{[2]}Syracuse University,

\begin{abstract}

Graph Convolutional Networks (GCNs) are pivotal in extracting latent information from graph data across various domains, yet their acceleration on mainstream GPUs is challenged by workload imbalance and memory access irregularity. To address these challenges, we present \ourframework, a GPU accelerator architecture for GCNs. The design of \ourframework encompasses: (i) a lightweight degree sorting stage to group nodes with similar degree; (ii) a block-level partition strategy that dynamically adjusts warp workload sizes, enhancing shared memory locality and workload balance, and reducing metadata overhead compared to designs like GNNAdvisor; (iii) a combined warp strategy that improves memory coalescing and computational parallelism in the column dimension of dense matrices.

Utilizing these principles, we formulate a kernel for SpMM in GCNs that employs block-level partitioning and combined warp strategy. This approach augments performance and multi-level memory efficiency and optimizes memory bandwidth by exploiting memory coalescing and alignment. Evaluation of \ourframework across 18 benchmark graphs reveals that it outperforms cuSPARSE, GNNAdvisor, and graph-BLAST by factors of $1.17\times$, $1.86\times$, and $2.94\times$ respectively. The results underscore \ourframework as an effective solution for enhancing GCN computational efficiency. The implementation can be found on Github\footnote{\url{https://github.com/xiexi1990/ICCAD-Accel-GNN}}. 

\end{abstract}


\footnotetext[2]{H. Fang is now affiliated with Synopsys, Mountain View, CA. }

\section{Introduction}

% Graph convolution networks (GCNs)~\cite{kipf2016semi, liu2020graphsage} are commonly used for link prediction~\cite{zhang2018link, hu2020open}, node classification~\cite{kipf2016semi, chen2020simple, hu2020open} generative drug discovery~\cite{bongini2021molecular}, and recommendation systems \cite{ying2018graph}. In the link prediction tasks, GNN tries to predict the possible unseen link connections between nodes. For node classification tasks, GNNs predict the target node classes using their output embeddings. 

Graph Convolutional Networks (GCNs)~\cite{kipf2016semi, liu2020graphsage} are a type of Graph Neural Networks (GNNs) that has drawn tremendous attention in the past years due to their unique ability to extract latent information from graph data~\cite{hu2020open}. Practical applications of GCNs include prediction of cascading power-grid failure \cite{liu2020guiding}, traffic forecasting~\cite{jiang2022graph}, recommendation systems~\cite{wu2020graph}, and drug discovery~\cite{bongini2021molecular}. The deployment of GCNs in these applications typically poses strict constraints on latency and throughput.

When designing GNN accelerators, GPU platforms have emerged as the mainstream choice.  Existing GCN acceleration designs mainly process a moderately sparse graph feature matrix ($X$) multiplication with a dense and small weight matrix ($W$), and then multiply the output with the highly sparse and irregular adjacency matrix ($A$). 
% They exhibit two main challenges: workload imbalance and memory access irregularity.
They exhibit two main challenges: workload imbalance and data locality.
The power-law distribution prevalent in the $A$ matrix of a graph often leads to significant sparsity and irregularity~\cite{geng2020awb}, which brings challenges to workload mapping for existing hardware platforms. Conventional workload partition methods ~\cite{wang2021gnnadvisor} for $A\cdot X$ operation may result in workload imbalances across various warps. Consequently, simple row-wise workload allocation of $A$ workload could trigger idleness in certain threads, inhibiting overall performance. Efficient SpMM algorithms for $A\cdot X$ necessitate the effective management of parallelism across both the sparse matrix rows or columns and the resulting dense matrix accumulation to optimally distribute workload, ideally at the warp level.

Contemporary GPUs display a mutli-level memory hierarchy~\cite{jia2018dissecting}, and SpMM operations employ memory-efficient formats like Compressed Sparse Row (CSR) or Compressed Sparse Column (CSC), resulting in irregular data structures and non-coalesced memory accesses.
State-of-the-art (SOTA) approaches, including GNNAdvisor~\cite{wang2021gnnadvisor}, Graph-BLAST~\cite{yang2018implementing}, and cuSPARSE~\cite{naumov2010cusparse}, have sought to optimize SpMM performance. 
% but exhibit limitations. 
GNNAdvisor's use of non-zero groups (NG)~\cite{wang2021gnnadvisor} enhances workload mapping but can result in warp-level workload imbalance and resource underutilization on graphs with power-law non-zero distribution. Graph-BLAST~\cite{yang2018implementing}, though supporting diverse graph operations and improving memory coalescing, lacks efficiency in SpMM, particularly in dense matrix column dimension traversal. CuSPARSE~\cite{naumov2010cusparse}, a strong baseline for SpMM kernels, restricts further insight due to its closed-source nature. 
% These challenges highlight the need for continued refinement in SpMM techniques, recognizing that existing solutions, while valuable, have specific constraints that may hinder optimal performance.
Overall, these approaches encounter performance bottlenecks in addressing workload balance~\cite{wang2021gnnadvisor}, efficiency in dense dimension traversal~\cite{yang2018implementing}, and limitations in extensibility or insight due to closed-source development~\cite{naumov2010cusparse}.
% \cd{summarize their limitations here. Focus on data locality.}
% In this research, we present \ourframework, an advanced open-source GPU kernel design GCNs, which surpasses SOTA methods and libraries, including GNNAdvisor~\cite{wang2021gnnadvisor}, graph-BLAST \cite{yang2018implementing} and cuSPARSE~\cite{naumov2010cusparse}. \ourframework integrates degree sorting, block-level partition and combined warp techniques to optimize data locality, multi-level memory efficiency, workload assignment and memory access coalescing. 
% We design the CUDA kernel that fully capitalizes on the advantages of these techniques. Our contributions are outlined as follows:

In this research, we introduce \ourframework, an open-source GPU kernel design for GCNs that outperforms SOTA methods and libraries, including GNNAdvisor~\cite{wang2021gnnadvisor}, graph-BLAST~\cite{yang2018implementing}, and cuSPARSE~\cite{naumov2010cusparse}. \ourframework aims to enhance various computational aspects such as data locality, multi-level memory efficiency, workload assignment, and memory access coalescing through the integration of degree sorting, block-level partition, and combined warp techniques. The core contributions of this work are encapsulated in the design of the CUDA kernel that leverages the aforementioned techniques and are outlined as follows:

\begin{itemize}
\item A degree sorting based preprocessing step with $\mathcal{O}(n)$ complexity is proposed. This lightweight step aims to enhance data locality and facilitate workload mapping by grouping rows with identical degrees together. 

%Degree sorting ensures an $\mathcal{O}(n)$ complexity when combining with a linear time-complexity sorting algorithm.
% \cd{add the direct advantage.}
% .Rows with a higher degree in the graph cause workload imbalance. Degree sorting groups these rows together,
% enhancing data locality and countering the workload imbalance. 
% By employing a linear time-complexity sorting algorithm, degree sorting has an $\mathcal{O}(n)$ complexity. It boasts a speed advantage that can be implemented on-the-fly. 

\item
A block-level partition scheme is developed to dynamically adjust warp workload sizes across different blocks. Compared to SOTA designs like GNNAdvisor~\cite{wang2021gnnadvisor} which use a fixed workload size assignment, the dynamical allocation creates a more balanced workload among warps. We further customize a metadata format for block-level partition, 
% The block-level partition further reduces overhead by 
enabling all warps within a block to share a single metadata for workload mapping. As such, this strategy enhances workload balance and shared memory reuse efficiency.

\item
A combined warp strategy is formulated to maximize thread-address continuity in the traversal and processing of dense matrix's column dimension. This approach furthers the cause of column dimension memory coalescing and computational parallelism, leading to a more efficient execution.

% synergized with shared memory alignment, to optimize the execution efficiency further. This approach combines both techniques in a cohesive framework, allowing for improved performance and memory bandwidth utilization.

\end{itemize}

We conduct evaluation of \ourframework on 18 benchmark graphs and observe a significant performance boost. The average improvements are $1.17 \times$ over cuSPARSE~\cite{naumov2010cusparse}, $1.86\times$ over GNNAdvisor~\cite{wang2021gnnadvisor}, and $2.94\times$ over graph-BLAST~\cite{yang2018implementing}.

\section{Preliminary and Related Work}
% \section{Challenges and Existing Work}
% \cite{zhang2019efficient}
% \textcolor{red}{draw a figure or GNN to GPU mapping, to illustrate each challenges}

% Challenges and solutions: \\
% 1. GNN basics (done)

% 2 (challenges). For GPU: 1.1 Computation, power lower distribution, hard to partition workload,  1.2 memory access

% 3. To solve 1.1 computation, how to propose new workload partiiton, existing partition. Different format? distribute workload, example: C2SR for channle-wise distribution

% 4. To solve 1.2 computation, memory optimization method. main contribution: Shared memory and alignment. \\
% Global memory? not main contribution, don't discuss much

\subsection{Graph Convolution Network}

\begin{figure}[t]
\centering
    \includegraphics[
    clip, 
    width =0.9 \linewidth]{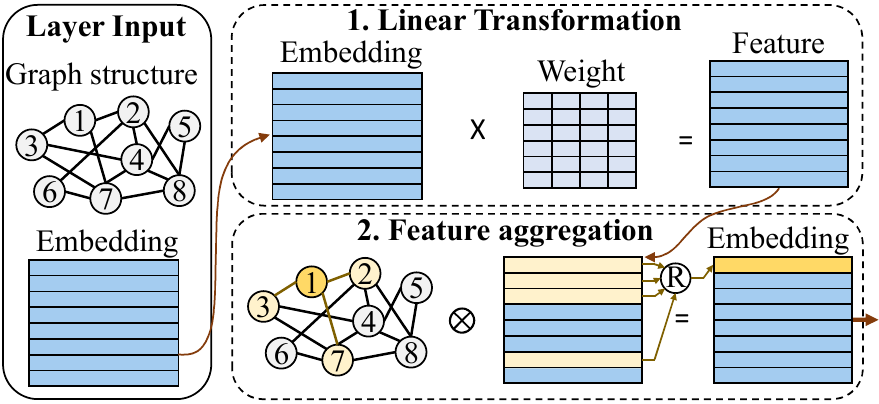} 
     \caption{Computational workflow of single GCNConv layer.}
    \label{fig:GCN_structure} 
    \vspace{-0.2in}
\end{figure}

Graph Convolutional Networks (GCNs)~\cite{kipf2016semi}, comprised of GCNConv layers, undergo two primary stages: linear transformation and feature aggregation, as illustrated in Fig.~\ref{fig:GCN_structure}. Given a graph \( G = ( \mathcal{V}, \mathcal{E}, A ) \), where the adjacency matrix \( A \) represents the existence of edges between nodes, the forward propagation in the \( l \)-th GCNConv layer can be decoupled into: (1) linear transformation, \( Y^l = X^l W^l \), and (2) feature aggregation, \( X^{l+1} = \sigma(A' Y^l) \). Here, \( X^l \) is the feature embedding matrix, \( W^l \) denotes the weight matrix, and \( A' \) is the normalized adjacency matrix. The activation function, typically an element-wise ReLU, calculates the feature embedding matrix output.

GCN variants like GraphSAGE~\cite{hamilton2017inductive} and Graph Isomorphism Network (GIN)~\cite{hou2019measuring} maintain similar structures but with distinct aggregation functions, upholding the same forward propagation model as traditional GCNs. The efficiency of GCNs is contingent on the feature aggregation stage~\cite{geng2020awb, geng2021gcn}, chiefly executed as sparse matrix multiplication (SpMM) between the adjacency list \( A' \) and embeddings \( Y^l \). This ultra-irregular operation is typical in GCNConv layers, and given the significant role of SPMM in GCN, its acceleration is essential for boosting GCN performance.

% In this paper, we propose GE-SpMM (General-purpose SpMM), an efficient SpMM kernel based on the CSR format that can accelerate GNN workloads on GPUs. GE-SpMM can seamlessly integrate with various GNN algorithms without the need for data conversion overhead.

% SpMM is a fundamental operator in GNN, performing multiplication between a sparse matrix and a dense matrix. In GCN, nodes and edges can be represented as matrices, and SpMM can be used to compute the product between these matrices. For example, in Graph Convolutional Networks (GCN), SpMM is used to compute the product between the adjacency matrix and feature matrix.

% As a large number of SpMM operations are required in GNN, accelerating SpMM is crucial for improving GNN performance. 
% % In this paper, we propose GE-SpMM (General-purpose SpMM), an efficient SpMM kernel based on the CSR format that can accelerate GNN workloads on GPUs. GE-SpMM can seamlessly integrate with various GNN algorithms without the need for data conversion overhead.
% Therefore, it can be said that SpMM is an indispensable operator in GNN, and accelerating SpMM is crucial for improving GNN performance.

\subsection{SpMM Acceleration}
% \subsection{Existing works and Challenges}
% \subsection{SPMM Algorithm Designs Specific for GPU}

% \textcolor{red}{@xiexi, summarize existing GPU works}

GraphBLAST~\cite{yang2018design, yang2018implementing}, GE-SpMM~\cite{huang2020ge-spmm}, and Jiang et al.~\cite{jiang2020novel} present distinct methods for enhancing sparse matrix-matrix (SpMM) multiplication on GPUs. GraphBLAST adopts a block-based SpMM algorithm utilizing instruction-level parallelism to minimize latency and combines it with a "row-splitting" memory access pattern and "static scheduling" load balancing for efficient matrix access and adaptive thread allocation. Conversely, GE-SpMM offers an effective CSR-based SpMM kernel for generalized integration with Graph Neural Network (GNN) algorithms, eliminating data conversion overhead. Jiang et al. introduce a row-reordering technique to improve SpMM performance across hardware platforms and investigate its applicability for diverse parameter settings.

GNNAdvisor~\cite{wang2021gnnadvisor} introduces an adaptive runtime system designed for GNN acceleration on GPUs, incorporating specialized memory optimizations, community-aware node renumbering, and warp-aware memory customization. Notably, SpMM is utilized in Deep Graph Library (DGL) for sum-reduced aggregation. While these methods represent significant advancements, limitations persist in practical applications, primarily due to right-multiply matrix dimensional constraints~\cite{yang2018design, yang2018implementing, huang2020ge-spmm, jiang2020novel}. Jiang et al.'s row-reordering method is challenging to implement on-the-fly, and GNNAdvisor relies on shared memory caching for performance improvements, which does not entirely benefit from memory alignment and varies with the right-hand matrix dimensions. 
% MergePath~\cite{shan2023mergepath} employs binary search algorithm to conduct fine-grained balanced workload mapping, 
% but it still encounters the issue of lower processing efficiency when handling high degree rows in the graphs with power-law distribution.

%fixed cost parameter 

In summary, two primary challenges within the SpMM domain require optimization: computational workload unbalance and memory access irregularity.

% Sparse Matrix-Dense Matrix Multiplication (SpMM) algorithm design faces challenges in both computation and memory aspects when implemented on parallel architectures like GPUs. Here, we discuss these challenges:

%Due to the power-law distributed irregular non-zero elements in the graph adjacent matrix~\cite{geng2020awb}, there might be varying amounts of workload been allocated for different threads or thread blocks, leading to load imbalance. 

\section{\ourframework Framework}

% \begin{itemize}

% \item 
% The graph adjacent matrix is normally power-law distributed~\cite{geng2020awb} and extremely sparse and irregular. The nave workload partition~\cite{wang2021gnnadvisor} may lead to load imbalance across different warps and blocks. An example of Collab~\cite{graph_collab} degree distribution is shown in Fig.~\ref{fig:collab_histogram}, the nodes degree can be as much as 671, which is 66 times higher than the average degree. As such, simple row-wise workload allocation may result in some threads being idle while others are still busy processing, reducing overall performance. On the other hand, SpMM requires the exploitation of multiple levels of parallelism to efficiently utilize GPU resources. This includes parallelism across the rows or columns of the sparse matrix and within the computation of the resulting dense matrix elements. Designing an algorithm that effectively manages this parallelism can be challenging. Consequently, an SpMM algorithm must distribute the workload reasonably and efficiently, typically down to the level of individual warps.

\subsection{Motivation}
% We present two pivotal observations that serve as the foundation for our \ourframework. 
\subsubsection{\textbf{Workload Allocation Matters}}
\begin{figure}[htbp]
    \centering
    \includegraphics[
    clip, 
    width =0.75 \linewidth]{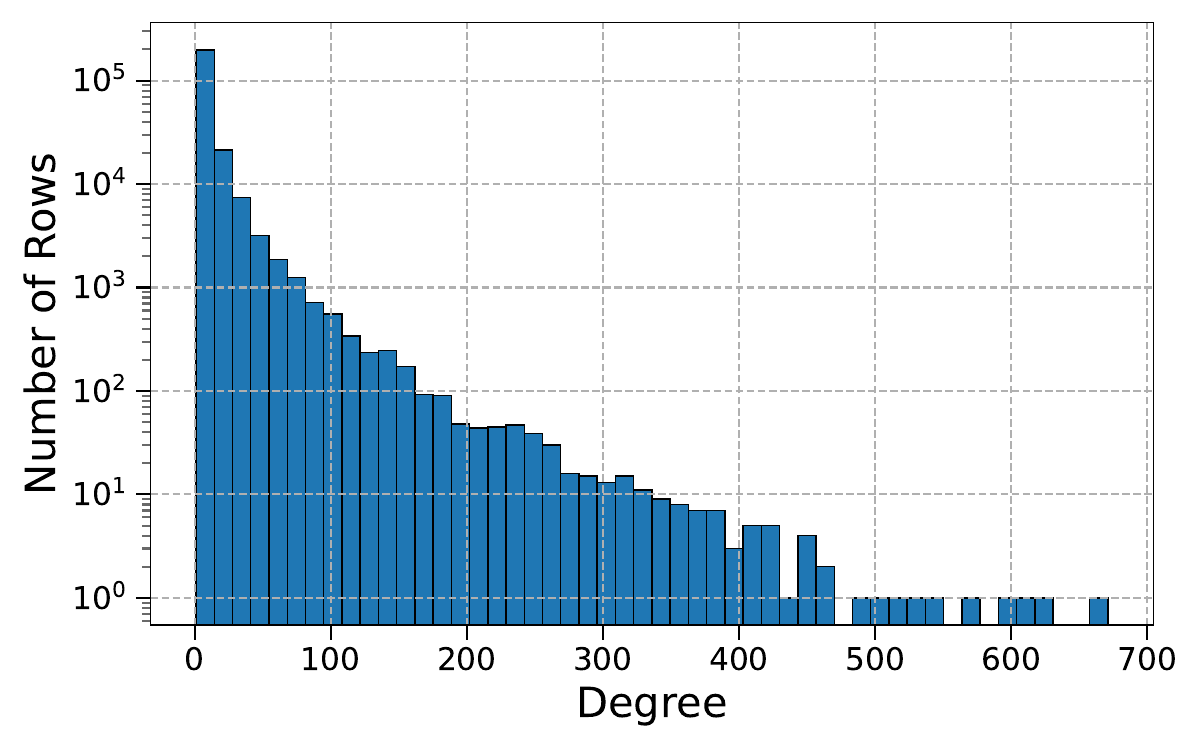} 
     \caption{
     A histogram for the row degree distribution of Collab. 
     }
    \label{fig:collab_histogram} 
\end{figure}

The adjacency matrix of a graph, often exhibiting a power-law distribution and extreme sparsity~\cite{geng2020awb}, can lead to load imbalance across warps and blocks with naive workload partitioning~\cite{wang2021gnnadvisor}. As shown in the Collab graph degree distribution~\cite{graph_collab}, nodes can have degrees up to 66 times greater than the average, as presented in Fig.~\ref{fig:collab_histogram}. This discrepancy may cause uneven workload allocation, resulting in idle threads and worse performance.

% The adjacency matrix of a graph is often characterized by a power-law distribution~\cite{geng2020awb}, resulting in extreme sparsity and irregularity. Naive workload partitioning~\cite{wang2021gnnadvisor} may cause load imbalance across different warps and blocks. For instance, the Collab graph degree distribution~\cite{graph_collab} presented in Fig.~\ref{fig:collab_histogram} demonstrates that nodes can have degrees as high as 671, which is 66 times greater than the average degree. Consequently, simple row-wise workload allocation may cause some threads to become idle while others are still processing, adversely impacting overall performance. 

Moreover, efficient utilization of GPU resources in SpMM requires exploiting multiple levels of parallelism, encompassing parallelism across rows or columns of the sparse matrix and the computation of resulting dense matrix elements. Designing an algorithm that effectively manages this parallelism can be challenging. Therefore, an SpMM algorithm must distribute the workload judiciously and efficiently, ideally down to the level of individual warps.

\subsubsection{\textbf{Memory Access Patterns Matters}} Modern GPUs exhibit a three-level memory hierarchy~\cite{jia2018dissecting}: global DDR memory, cache (L1 and L2), and shared memory. In SpMM operations, sparse matrices use efficient representations such as CSR or CSC, leading to irregular data structures and non-coalesced memory accesses. These access patterns can result in increased latency~\cite{geng2020awb, jiang2020novel}, memory bank conflicts, and reduced memory bandwidth utilization. To optimize SpMM performance, all levels of the GPU memory hierarchy should be considered. We summarize two challenges as follows: 

\textit{a) Shared Memory Mapping Challenge.} Efficient use of shared memory can help minimize global memory access latency~\cite{jia2018dissecting}. However, due to the irregular structure of sparse matrices, effectively using shared memory for SpMM can be challenging. Loading data into shared memory might involve complex indexing and synchronization, which can increase overhead and diminish performance benefits.

\textit{b) Cache Locality Challenge.} The irregular memory access patterns in SpMM can lead to inefficient cache utilization~\cite{zhao2019adaptive}, causing cache thrashing and increased memory access latency. Designing algorithms that maximize cache utilization can be challenging due to the non-uniform distribution of non-zero elements in the sparse matrix.

To address these challenges, various optimization techniques can be applied, including selecting appropriate sparse matrix formats \cite{ortega2013fastspmm, jiang2020novel}, exploiting multiple levels of parallelism \cite{yang2018design}, implementing coalesced memory access patterns \cite{huang2020ge-spmm}, efficiently using shared memory \cite{wang2021gnnadvisor}, and employing dynamic load balancing techniques \cite{geng2020awb}. However, in spite of these optimizations, SpMM often remains a complex task in terms of both computational and memory aspects.

\subsection{\ourframework Preliminary}
Sparse Matrix Representation and Reordering significantly influence SpMM algorithm performance, with proper techniques enhancing memory access, minimizing complexity, and optimizing performance. Research has developed reordering input data and new sparse matrix representations, such as ELLPACK-R in FastSpMM~\cite{ortega2013fastspmm}, SELLP in MAGMA~\cite{anzt2015accelerating}, register blocking in OSKI~\cite{vuduc2005oski}, and Compressed Sparse Blocks (CSB)~\cite{aktulga2014optimizing}. These implementations have yielded performance gains. RS-SpMM~\cite{hong2018efficient} introduced a format for better SpMM data locality, albeit with limitations.

% I-GCN~\cite{geng2021gcn} assessed graph reordering methods including HATS~\cite{8574527}, SlashBurn~\cite{lim2014slashburn}, and Rabbit~\cite{arai2016rabbit}, each offering different trade-offs in efficiency, complexity, parallelization, and suitability for specific frameworks, highlighting a continued balance between enhancement and overhead in sparse matrix operations.

I-GCN~\cite{geng2021gcn} evaluated graph reordering techniques including HATS~\cite{8574527}, SlashBurn~\cite{lim2014slashburn}, and Rabbit~\cite{arai2016rabbit}, each with their own strengths and limitations. HATS~\cite{8574527} enhances cache hierarchy efficiency, whereas SlashBurn~\cite{lim2014slashburn} clusters non-zeros effectively but requires complex, non-parallelizable logic. Rabbit reordering~\cite{arai2016rabbit} outperforms other approaches in terms of data locality, parallelization ease, and performance but is unsuitable for \emph{\ourframework} due to its processing overhead compared to GCN inference.

% \textbf{Shared Memory Utilization and Alignment}
% Efficient shared memory utilization and alignment are crucial for enhancing GPU performance. Coalesced memory access~\cite{cuda_toolkit, kirk2016programming} helps minimize memory transactions, improving performance. Memory access alignment becomes challenging when writing final results to global memory. Padding the shared memory array to the closest multiple of 32 allows for alignment optimizations when managing intermediate SpMM results~\cite{huang2020ge-spmm, yang2018design, yang2018implementing}.

\textbf{Shared Memory Utilization and Alignment}
Efficient alignment and utilization of shared memory are vital for optimizing GPU performance. Coalesced memory access~\cite{cuda_toolkit, kirk2016programming} enhances this efficiency but aligning access when writing to global memory can be challenging. Padding the shared memory array to the nearest multiple of 32 optimizes alignment when handling intermediate SpMM results~\cite{huang2020ge-spmm, yang2018design, yang2018implementing}. Though optimizing global memory alignment is complex, shared memory offers opportunities for alignment improvement, thus potentially enhancing SpMM performance, given proper padding and indexing management. Some previous works, such as GraphBLAST~\cite{yang2018design, yang2018implementing}, GE-SpMM~\cite{huang2020ge-spmm}, Jiang et al.~\cite{jiang2020novel}, and GNNAdvisor~\cite{wang2021gnnadvisor}, have faced alignment inefficiencies in corner cases, leading to suboptimal performance.

Although optimizing memory access alignment for global memory is challenging, using shared memory for intermediate results offers opportunities for alignment optimization, potentially improving SpMM performance. However, proper management of padding and indexing is necessary to ensure accurate results and avoid memory access conflicts.

Certain prior works, including GraphBLAST~\cite{yang2018design, yang2018implementing}, GE-SpMM~\cite{huang2020ge-spmm}, Jiang et al.~\cite{jiang2020novel}, GNNAdvisor~\cite{wang2021gnnadvisor}, and MergePath-SpMM~\cite{shan2023mergepath}, have encountered inefficient alignment in corner cases, resulting in suboptimal memory system performance.

\subsection{\ourframework Preprocessing}
% \textcolor{red}{add some data support for sorting algorithm time}
Our \textit{\ourframework} design incorporates  two key preprocessing steps—degree sorting and block-level partitioning—to enable efficient parallel processing of sparse matrices.

% \textbf{Degree Sorting.} Sorting a CSR-formatted sparse matrix by degree comprises the following steps: computing row degrees using row pointer arrays, sorting rows based on their calculated degrees, and updating row pointer arrays to reflect the new row order. By employing a counting sort algorithm~\cite{sun2009count} with a linear time complexity of $O(n)$, the operation's overall time complexity can be optimized. Degree sorting exhibits a substantially lower complexity than other algorithms and can be executed on the fly, demonstrating higher efficiency.

%Optimizing Memory Utilization: Unveiling the Benefits of Block-Level Partitioning over Warp-Level Partitioning in GPU Computing.
\begin{figure}[ht!]
    \includegraphics[
    clip, 
    width =0.99 \linewidth]{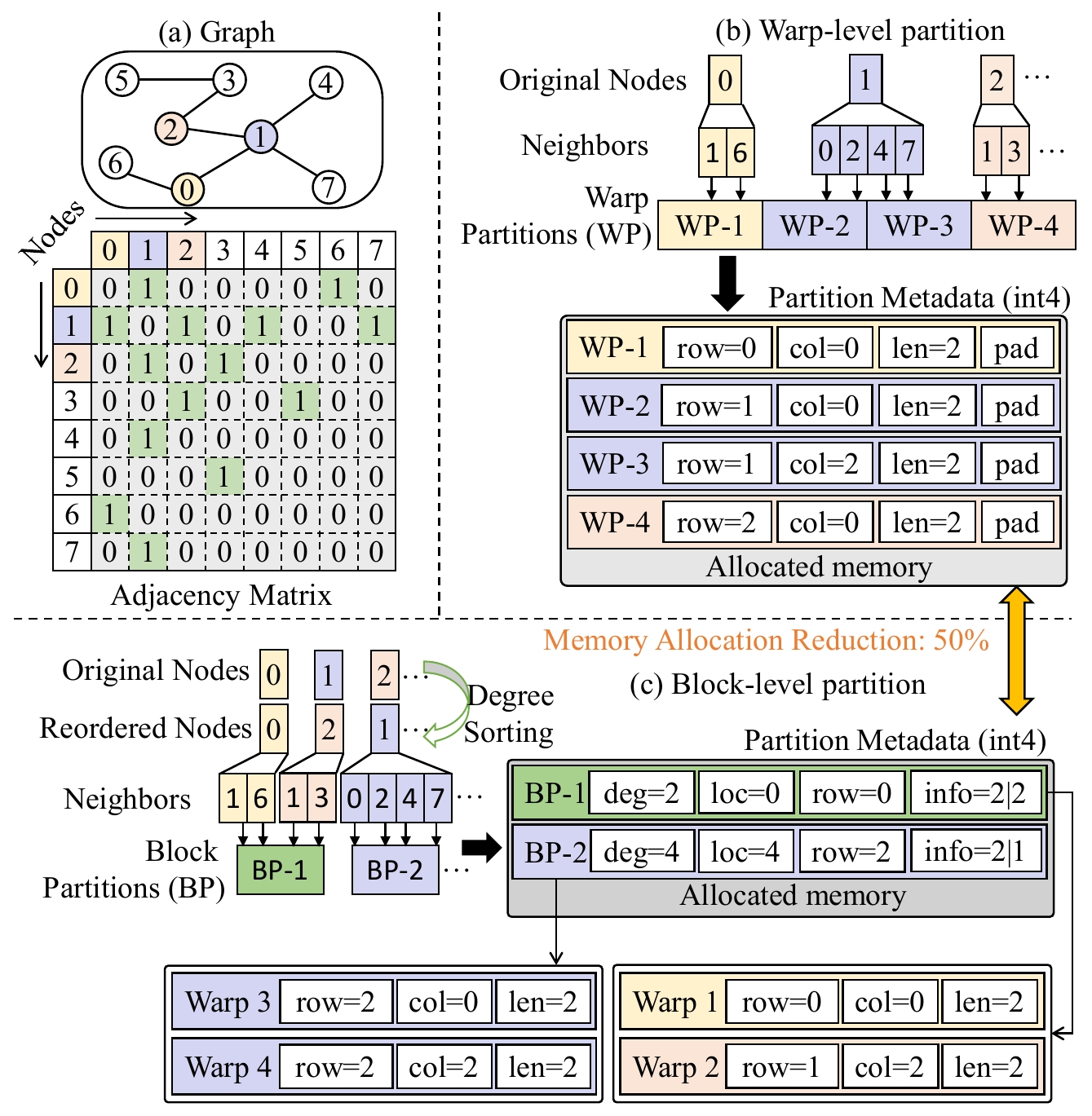} 
     \caption{Metadata generation process: (a) original graph structure. (b) Metadata of warp-level partition. (c) Metadata of block-level partition. }
    \label{fig:block_level_partition} 
\end{figure}

\textbf{Degree Sorting.} Degree sorting serves as a preliminary step for block-level partitioning. Sorting sparse matrix degree in a CSR-formatted sparse matrix requires the following steps: (1) computing each row's degree using the row pointer array, which has a time complexity of $\mathcal{O}(n)$ when employing count sort \cite{sun2009count} or radix sort \cite{merrill2011high}, with n indicating the number of rows; (2) applying a stable sorting algorithm to sort rows based on the degrees; and (3) updating the row pointer array to reflect the new row order, with a time complexity of $\mathcal{O}(n)$. The dominant time complexity of this operation arises from applying the stable sorting algorithm. Nevertheless, employing count sort, a linear time-complexity algorithm, can optimize the overall time complexity to $\mathcal{O}(n)$. This lower time complexity enhances efficiency compared to alternative algorithms and allows on-the-fly execution.

\begin{figure*}[htbp]
    \includegraphics[
    clip, 
    width =0.99 \linewidth]{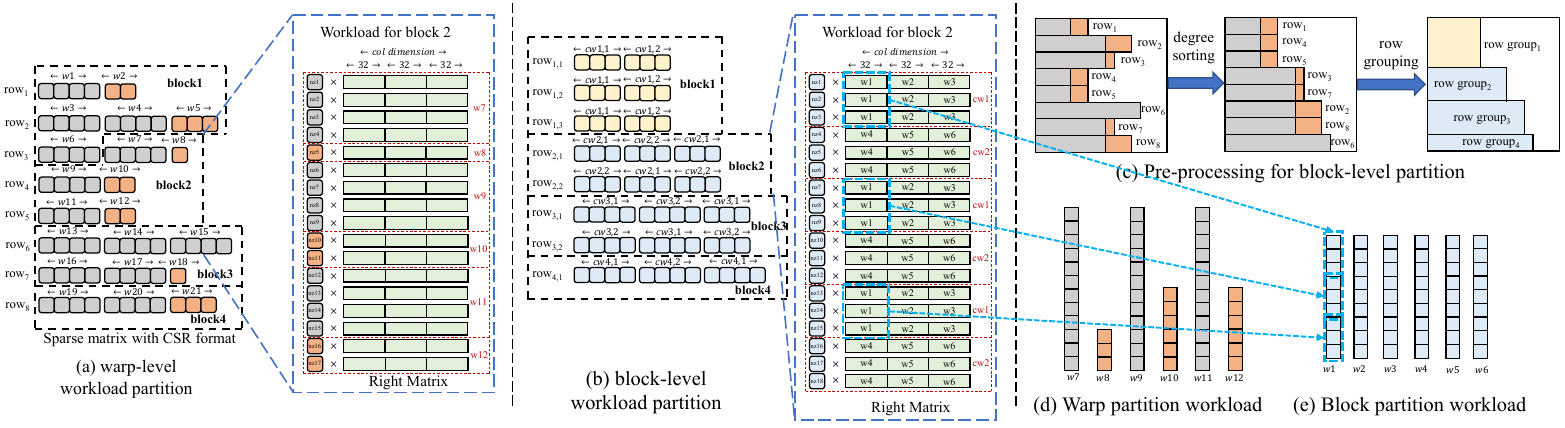} 
     \caption
     {
     Illustration of hardware mapping strategy: (a) the warp-level workload partitioning with a single warp traversing the column dimension of the dense matrix. (b) The block-level workload partitioning (right) with the combined warp strategy handling the column dimension of the dense matrix. (c) Preprocessing steps for block-level partitioning strategy. (d) Workload distribution of warp-level partitioning. (e) Workload distribution of block-level partitioning.
     % The demonstration of two computational processes: (left) warp-level workload partitioning combined with a single warp traversing the column dimension of the dense matrix; (right) block-level workload partitioning combined with the processing of the dense matrix column dimension using a combined warp strategy. 
     %\textcolor{red}{1.How reduce meta data/better partition method can increase hardware performance 2. Why block partition help to reduce residual part? why residual part is important?}
     }
    \label{fig:block_vs_warp} 
\end{figure*}

% As illustrated in Fig.~\ref{fig:block_vs_warp}, unlike warp-level partitioning, which requires storing metadata for all warps, in block-level partitioning, all warps in each block share the same metadata, so the metadata that needs to be stored is only at the magnitude of the number of blocks. 
% The storage space requirement ratio between block-level and warp-level partitioning is:

% \begin{gather}
% \frac{S_{B}}{S_{W}} \approx \frac{1}{\text{Avg. Warps per Block}}
% \end{gather}

% This typically results in less than 10\% of the storage space needed for warp-level partitioning.
% For example, if $max\_block\_warps$ equals 12, only about 8\% of the memory remains in use for meta-data with block-level partitioning.

\begin{algorithm}[t]
\caption{Get partition patterns}
\label{alg:get_partition_patterns}
% \hspace{16pt}
% \comment{Initialization settings}
\begin{algorithmic}[1]
% \cd{add definitions of 12, 32.]}
% \STATE $max\_block\_warps \gets 12;max\_warp\_nzs \gets 32$;
\STATE $deg\_bound \gets max\_block\_warps \times max\_warp\_nzs$;
\STATE $factors \gets$ all factors of $max\_block\_warps$;
\STATE $i \gets 0; deg \gets 1;$\\
\WHILE{$deg < deg\_bound$}
    \IF{$factors[i] \times warp\_max\_nz \ge deg$}
        \STATE assign $block\_rows \gets \frac{max\_block\_warps}{factors[i]}$ to current block-partition pattern;
        \STATE assign $warp\_nzs \gets \lceil \frac{deg}{factor[i]} \rceil$ to current block-partition pattern;
        % \STATE $pattern[deg].block\_rows \gets$\\$max\_block\_warps \mathbin{/} factor[factor\_cnt];$
        % \STATE $pattern[deg].warp\_nzs \gets$\\ ceil $(deg \mathbin{/} factor[factor\_cnt]);$
        \STATE $deg$ ++;
    \ELSE
        \STATE $i$ ++; 
    \ENDIF
\ENDWHILE
\end{algorithmic}
\end{algorithm}

\begin{algorithm}[ht!]
\caption{Block-level partitioning}
\label{alg:block_level_partitioning}
% \hspace{16pt}
% \comment{Use patterns to do partitioning}
\begin{algorithmic}[1]

\FOR{each $deg$}
    \IF{$deg \leq deg\_bound$}
        \STATE $row\_remaining \gets$ total number of rows of $deg$\\
        \WHILE{$rows\_remaining$ \\ $\geq pattern[deg].block\_rows$ }
            \STATE $\{row$, $loc$, $deg$, $warp\_nzs | block\_rows \} \rightarrow$ current metadata;
            % \STATE fill current meta-data with current $row$, $loc$, $deg$ and current block-partition pattern's $warp\_nzs$ and $block\_rows$ as additional information;
            % \STATE $block\_row\_begin \gets current\_row$;\\
            % \STATE $block\_loc\_begin \gets current\_loc$;\\
            % \STATE $block\_deg \gets deg$;\\
            % \STATE $block\_info \gets (pattern[deg].warp\_nzs<< 16)$\\$ | pattern[deg].block\_rows$;\\
            \STATE $rows\_remaining \minuseq pattern[deg].block\_rows$;
        \ENDWHILE
        \STATE $\{row$, $loc$, $deg$, $warp\_nzs | rows\_remaining \} \rightarrow$ current metadata;
        % \STATE fill current meta-data with current $row$, $loc$, $deg$ and current block-partition pattern's $warp\_nzs$ with $rows\_remaining$ as additional information;
        % \STATE $block\_row\_begin \gets current\_row$;\\
        % \STATE $block\_loc\_begin \gets current\_loc$;\\
        % \STATE $block\_deg \gets deg$;\\
        % \STATE $block\_info \gets (pattern[deg].warp\_nzs << 16)$\\$ | rows\_remaining$;
             
    \ELSE
        \STATE $deg\_remaining \gets deg$

        \WHILE{$deg\_remaining$$\geq deg\_bound$ }
            \STATE $\{row$, $loc$, $deg$, $deg\_bound \} \rightarrow$ current metadata;
            % \STATE fill blocks' meta-data with:
            % \STATE $block\_row\_begin \gets current\_row$;\\
            % \STATE $block\_loc\_begin \gets current\_loc$;\\
            % \STATE $block\_deg \gets deg$;\\
            % \STATE $block\_info \gets deg\_bound$;\\
            \STATE $deg\_remaining \minuseq deg\_bound$;
        \ENDWHILE
        \STATE $\{row$, $loc$, $deg$, $deg\_remaining \} \rightarrow$ current metadata;
        % \STATE fill the residual block's meta-data with:
        % \STATE $block\_row\_begin \gets current\_row$;\\
        % \STATE $block\_loc\_begin \gets current\_loc$;\\
        % \STATE $block\_deg \gets deg$;\\
        % \STATE $block\_info \gets deg\_remaining$;
    
    \ENDIF
\ENDFOR

\end{algorithmic}
\end{algorithm}

\textbf{Block-level Partition.} Block-level partition serves as a highly effective method for optimizing workload distribution among warps, which is also within a time complexity of $\mathcal{O}(n)$. This approach not only enhances computational resource allocation but also notably reduces the meta-data size necessitated for the partitioning process.

The block-level partition algorithm consists of two parts.
Algorithm \ref{alg:get_partition_patterns} presents the first part, get partition patterns, which entails determining the maximum number of warps per block and the maximum number of non-zero elements that each warp can handle. 
Their product, referred to as $deg\_bound$, signifies the maximum number of non-zero elements manageable by a single block.

For rows with a degree (number of non-zero elements) less than or equal to $deg\_bound$, each block processes one or more rows. By enumerating every factor $factor_i$ from 1 to $max\_block\_warps$, $factor_i$ warps process rows with degrees not exceeding $factor_i \cdot max\_warp\_nzs$, while $\frac{max\_block\_warps}{factor_i}$ rows are allocated to one block. When a row's degree surpasses $deg\_bound$, non-zero elements are assigned across multiple blocks to maximize loading.

The second part is described by Algorithm \ref{alg:block_level_partitioning}. After traversing through all rows of the graph once, the block-level partition algorithm generates meta-data for each block, shared by all warps within the same block. To fully exploit modern GPU read and write bandwidth, which permits reading and writing 128 bits simultaneously, the meta-data consists of an array of int4 data structures with a length equal to the number of blocks. The meta-data encompasses four elements: the block's degree, starting row number, starting address, and additional 32 bits of information. When the block's degree does not exceed $deg\_bound$, the additional information is split into two 16-bit segments, one for the number of non-zero elements handled by each warp and the other for the number of rows handled by the block. If the block's degree is greater than $deg\_bound$, the additional information stores the number of non-zero elements assigned to the block.
Since the block-level partitioning algorithm can be completed with a single pass through the rows of the graph, its time complexity is also within $\mathcal{O}(n)$. Therefore, the combined time complexity of degree sorting and block-level partitioning is also within $\mathcal{O}(n)$. Moreover, both algorithms are straightforward and suitable for on-the-fly execution.

%, which is equivalent to degree sorting on each row of the adjacency matrix

\textbf{Metadata Format for Block-level Partition.}
Figure~\ref{fig:block_level_partition} illustrates a representative example contrasting the metadata formats of block-level partition and warp-level partition.

In the warp-level partition depicted in Fig.~\ref{fig:block_level_partition}(b), each warp manages at most 2 non-zero elements. For instance, \textit{WP-1} oversees all non-zero elements of $row_0$ starting at $col_0$, and its corresponding metadata is ${row=0, col=0, len=2}$. Metadata for \textit{WP-2}, \textit{WP-3}, and \textit{WP-4} are constructed analogously. The cumulative warp metadata amount to 96 bits, necessitating 32 bits of padding to align with the 128-bit memory bus size.

Fig.~\ref{fig:block_level_partition}(c) explicates the block-level partitioning process. Initially, degree sorting is applied to the graph nodes, resulting in an ordered sequence of $row_0$, $row_2$, and $row_1$ for $row_0$, $row_1$, and $row_2$, respectively. Subsequently, block-level partitioning is executed such that each warp manages at most two non-zero elements and each block encompasses two warps. Consequently, both $row_0$ and $row_2$ (each with a degree of 2) encompassing a total of 4 non-zero elements are governed by \textit{BP-1}, while $row_1$ falls under the jurisdiction of \textit{BP-2}.

The metadata for block-level partitioning is encapsulated within an int4 array. It comprises the degree of the rows overseen by the block, the position of the first non-zero element, the row number of the initial row, and ancillary details, encompassing the quantity of non-zero elements handled by each warp and the number of rows managed by the block (each represented by 16 bits). Therefore, the metadata for \textit{BP-1} is ${deg=2, loc=0, row=0, info=2|2}$, and the metadata for \textit{BP-2} is ${deg=4, loc=4, row=2, info=2|1}$. The metadata storage of block-level partitioning normalized to warp-level partitioning is:

{
\small
\begin{gather}
\frac{S_{B}}{S_{W}} \approx \frac{1}{\text{Avg. Warps per Block}}
\end{gather}
}

Block-level partitioning, in comparison to warp-level partitioning, exhibits significant storage efficiency, typically requiring less than 10\% of the storage space. For instance, with a parameter of $max\_block\_warps$ set to 12, the block-level partitioning strategy necessitates a mere 8\% of the metadata storage relative to the warp-level approach.

One salient advantage of this approach is that the workload allocation for each warp within a block can be directly deduced from the block-level partition's metadata. Consider \textit{BP-1}, encompassing \textit{Warp-1} and \textit{Warp-2}. Given that the starting row of \textit{BP-1} is 0, the degree ($deg$) is 2, the number of accountable rows is 2, and each warp manages 2 non-zero elements, the responsibility for $row_0$ and $row_1$ is assigned to \textit{Warp-1} and \textit{Warp-2}, with corresponding column values of 0 and 2, respectively. This logic extends to other warps, such as \textit{Warp-3} and \textit{Warp-4} within \textit{BP-2}, enabling a consistent and systematic workload allocation.

The efficacy of block-level partitioning is further elucidated in Fig.~\ref{fig:block_vs_warp}(e). For all rows with a degree less than or equal to $deg\_bound$, the block-level partitioning patterns ensure a uniform workload distribution within each block. This stands in stark contrast to the warp-level partitioning, which exhibits a differentiated and uneven workload distribution. Consequently, the block-level partition patterns mitigate the decreased utilization of issue slots often associated with higher warp inactivity rates when handling residual workloads. This efficient alignment with the underlying computational architecture enhances parallelism and, ultimately, execution efficiency.

\begin{table*}[htbp]
\centering
\caption{Graph datasets details}
\resizebox{0.99\textwidth}{!}{
\begin{tabular}{|c|c|c|c|c|c|c|c|c|}
\hline
Graph Name & \# Nodes  & \# Edges   & Graph Name      & \# Nodes  & \# Edges   & Graph Name & \# Nodes  & \# Edges    \\ \hline
am         & 881,680   & 5,668,682  & amazon0601      & 403,394   & 5,478,357  & Artist     & 50,515    & 1,638,396   \\ \hline
Arxiv      & 169,343   & 1,166,243  & Citation        & 2,927,963 & 30,387,995 & Collab     & 235,868   & 2,358,104   \\ \hline
com-amazon & 334,863   & 1,851,744  & OVCAR-8H        & 1,889,542 & 3,946,402  & PRODUCTS   & 2,449,029 & 123,718,280 \\ \hline
Pubmed     & 19,717    & 99,203     & PPA             & 576,289   & 42,463,862 & Reddit     & 232,965   & 114,615,891 \\ \hline
SW-620H    & 1,888,584 & 3,944,206  & TWITTER-Partial & 580,768   & 1,435,116  & wikikg2    & 2,500,604 & 16,109,182  \\ \hline
Yelp       & 716,847   & 13,954,819 & Yeast           & 1,710,902 & 3,636,546  & youtube    & 1,138,499 & 5,980,886   \\ \hline
\end{tabular}}
\label{tab:graph_details}
\end{table*}

\begin{figure*}[htbp]
    \includegraphics[
    clip, 
    width = 1 \linewidth]{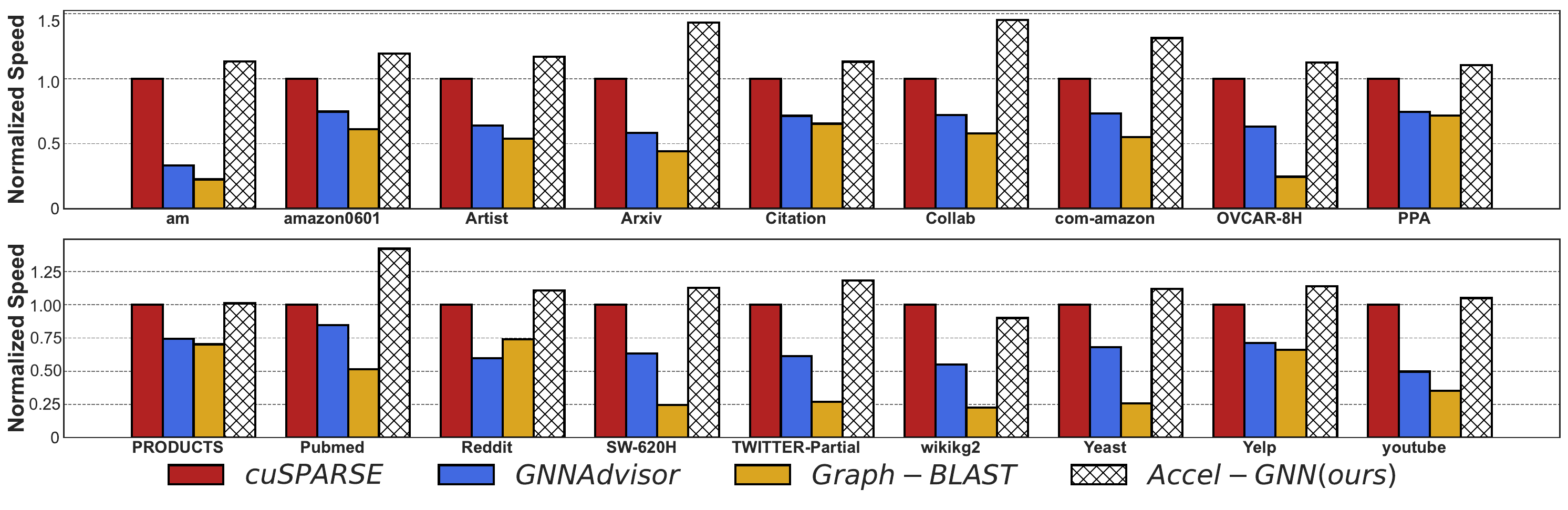} 
     \caption{
     Overall kernel performance comparison for cuSPARSE, GNNAdvisor, Graph-BLAST, and \ourframework (ours) on each graph. The speedup is normalized to cuSPARSE. 
     }
     % \caption{A comparison of normalized speeds between four methods such as: cuSparse, GNNAdvisor, Graph-BLAST, and Accel-GNN (ours) on a diverse range of benchmark graph datasets.}
    \label{fig:normalized_speed} 
\end{figure*}

\subsection{\ourframework Mapping}
\textbf{Combined Warp for Block-Warp Mapping.} 
% \textcolor{red}{Combined warp is a highly efficient organizational strategy for handling dense matrix dimensions, with the exploitation of memory coalescing and alignment to further optimize memory bandwidth.}
The combined warp approach represents an exceptionally efficient organizational strategy for addressing dense matrix dimensions. By leveraging memory coalescing and alignment, this method effectively optimizes memory bandwidth utilization, contributing to enhanced performance in the context of GPU acceleration.

SPMM is distinguished from Sparse Matrix-Vector Multiplication (SpMV) by its essential aspect of column dimension traversal of the right matrix. The traversal method for the right matrix significantly impacts performance, as SpMM is typically memory-bound.

When the column dimension of the right matrix surpasses the number of threads within a warp (usually 32), a single warp cannot accommodate the column dimension workload, necessitating traversal. Previous works, such as GNNAdvisor \cite{wang2021gnnadvisor}, adopt the natural method of introducing an inner loop for this warp. However, this approach introduces instruction-level branching and jumps, which, combined with memory positions mapped according to non-zero element positions, may fragment memory coalescing in the column dimension and lead to reduced efficiency.

To address these issues, we propose a combined warp execution approach that combines several consecutive warps, treating them as a single combined warp tasked with handling the entire column dimension workload. This method results in continuous memory position access within a combined warp's column dimension, improving cache hit rate and memory coalescing.

The implementation strategy for the combined warp unfolds as follows: First, compute the column dimension divided by 32 and round up. This produces the number of warps c within the combined warp, which defines $round\_dim = c \times 32$. The kernel computation's outermost loop includes an additional loop ranging from 0 to $c-1$, incrementing the thread id by the block dimension with each iteration to produce a new thread id. The warp id for the combined warp is then determined by dividing this thread id by $round\_dim$, and the division's remainder provides the lane id. Threads featuring a lane id greater than or equal to the right-hand matrix's column dimension are truncated, and the combined warp supplants the single warp for workload execution.

% We give a example of combined warp strategy shown in Figure~\ref{fig:block_vs_warp}(b). In the example, the dense matrix has a column dimension of 96, the combined warp strategy groups $w_1$ to $w_3$ together to form $cw_1$, and similarly, $w_4$ to $w_6$ are combined to form $cw_2$. $cw_1$ and $cw_2$ are responsible for all the workloads of $NZ_1$ to $NZ_3$ and $NZ_4$ to $NZ_6$ respectively. Subsequent workloads for other $NZ$s are sequentially assigned to $cw_1$ and $cw_2$. In contrast to the approach shown in Fig.~\ref{fig:block_vs_warp}(a) (adpoted in GNNAdvisor~\cite{wang2021gnnadvisor}) where a single warp loop through the column dimension to process the workload of single workload group, the combined warp strategy accesses the memory addresses in one row with continuous thread IDs, thereby maximizing the memory coalescing and computational parallelism.

An illustration of the combined warp strategy is provided in Fig.~\ref{fig:block_vs_warp}(b), elucidating its contrast to the approach presented in Fig.~\ref{fig:block_vs_warp}(a), as adopted in GNNAdvisor\cite{wang2021gnnadvisor}. In this example, a dense matrix with a column dimension of 96 is considered. The combined warp strategy groups warps $w_1$ to $w_3$ together to form a combined warp $cw_1$, and likewise, $w_4$ to $w_6$ are amalgamated to form $cw_2$. These combined warps, $cw_1$ and $cw_2$, are delegated with the responsibility of handling all workloads of $NZ_1$ to $NZ_3$ and $NZ_4$ to $NZ_6$, respectively. Subsequent workloads for other non-zero groups ($NZ$s) are sequentially assigned to $cw_1$ and $cw_2$.

Contrary to the approach in Fig.~\ref{fig:block_vs_warp}(a), where a single warp loops through the column dimension to process the single workload group, the combined warp strategy endeavors to access memory addresses in one row using continuous thread IDs. This method thereby maximizes memory coalescing and computational parallelism, offering an advantageous approach for optimizing memory access patterns and improving execution efficiency.

\begin{figure*}[htbp]
    \includegraphics[
    clip, 
    width =1 \linewidth]{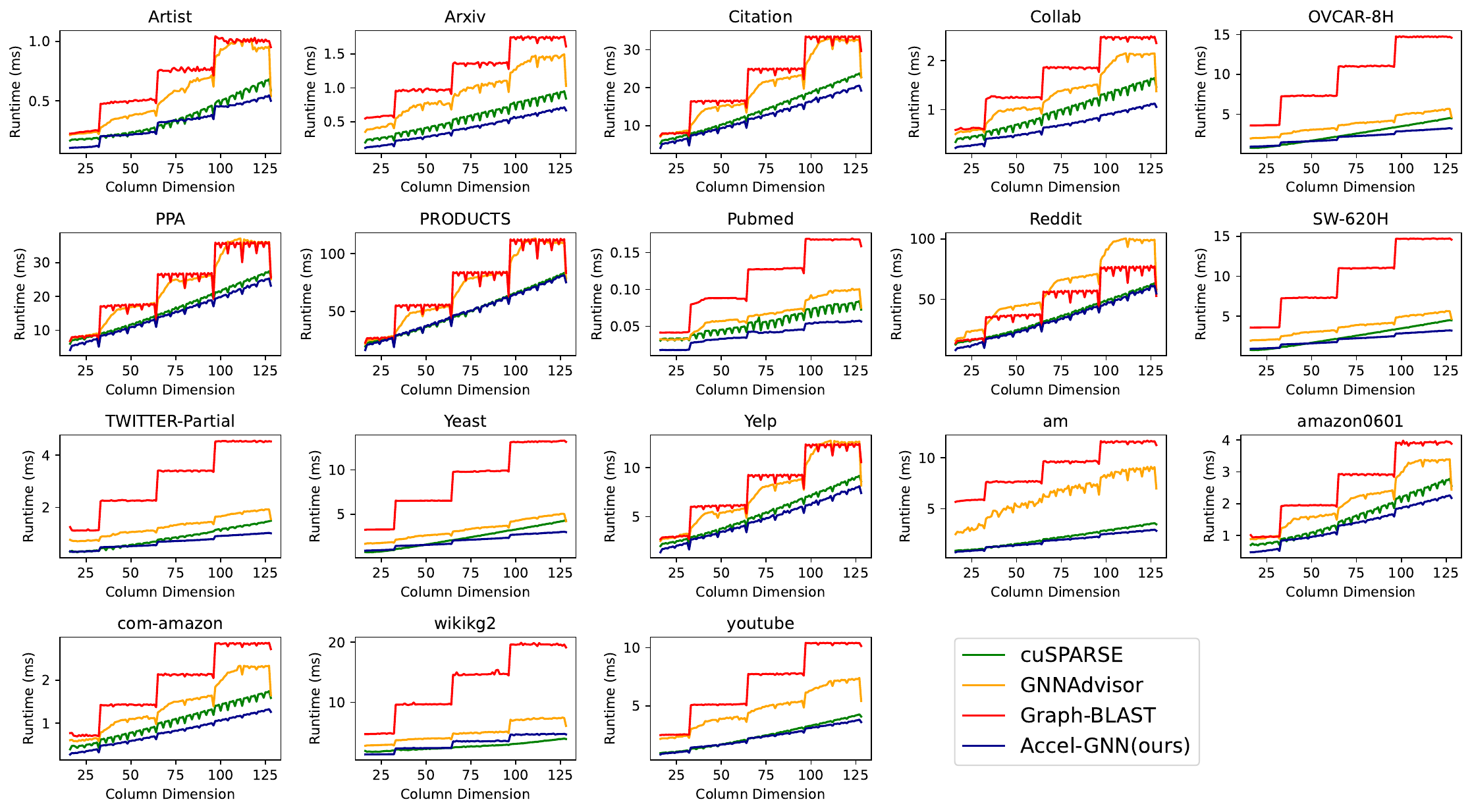} 
     \caption
     {The average SPMM kernel execution times for \ourframework (ours), GNNAdvisor, Graph-BLAST, and cuSPARSE have been tested on each graph, with the right-hand matrix's column dimensions ranging from 16 to 128.}
    \label{fig:runtime} 
\end{figure*}

\textbf{Summary and Further Enhancement}: We present an optimized computational approach that leverages the GPU's memory hierarchy and CUDA's $atomicAdd\_block$ feature. The proposed hierarchical warp computation strategy systematically accumulates partial results across three cache levels.

In the first cache level, independent parallel threads manage non-zero elements along the column dimension. The second cache level ensures atomicity among all combined warps handling identical row workloads within the same block. Utilizing CUDA's $atomicAdd\_block$ function, atomic operations within shared memory are enabled.

The third cache level addresses rows with degrees exceeding the $deg\_bound$ by concurrently executing partial results in multiple blocks and atomically accumulating them in global memory. The combined warp approach aligns shared memory, with higher kernel performance achieved when the column dimension is an integer multiple of 32.

Overall, this strategy efficiently utilizes memory hierarchy and atomic features, demonstrating significant performance gains and offering a pathway for future enhancements.

\section{Experimental Analysis}

\subsection{Experimental Framework}
The CUDA source code used in this study is compiled utilizing NVCC, version 12.0, and the execution is carried out on an Nvidia GeForce RTX 3090 platform equipped with Ubuntu 20.04. The experimental design involves the execution of SPMM with the left-hand sparse matrix from 18 benchmark graphs, specified in Table \ref{tab:graph_details}. The tests encompass scenarios where the column dimensions of the right-hand matrix vary from 16 to 128. 
%\cd{not include preprocessing results.}

The performance of our proposed kernel is gauged against benchmark techniques such as GNNAdvisor \cite{wang2021gnnadvisor}, Graph-BLAST \cite{yang2018design}, and the recent cuSPARSE, version 12.0. The latency measurements are conducted using the Nsight Compute~\cite{nsight_compute} tool. 
Note that the comparisons mainly focus on kernel execution time, excluding data transfer and preprocessing durations. Moreover, the impact on performance introduced by the adoption of block-level partition and combined warp strategy is evaluated.

The selection of graphs for this experiment involves popular benchmark datasets widely used in previous research \cite{hamilton2017inductive, kipf2017semi, xu2019how, kersting2016benchmark, fey2019fast, leskovec2014snap, wang2021gnnadvisor}. 
% such as amazon0601, Artist, com-amazon, OVCAR-8H, Pubmed, SW-620H, TWITTER-Partial, and Yeast, as well as datasets from the introductory high-performance computing lab at Tsinghua University \cite{tsinghua_spmm}, including am, Arxiv, Citation, Collab, PPA, PRODUCTS, Reddit, wikikg2, Yelp, and youtube. 
Table \ref{tab:graph_details} provides detailed parameters for each graph.
The range of graph sizes in the tests extends from a node count of 19,717 to 2,927,963, edge numbers ranging from 99,203 to 123,718,280, and densities spanning from $1.1 \times 10^{-6}$ to $2.1 \times 10^{-3}$. This diverse array of sizes and densities facilitates a comprehensive appraisal of the evaluated kernels' performance, demonstrating their scalability and proficiency under various circumstances.

\begin{table}[htbp]
\centering
\small
\caption{Impact of block-level partitioning and combined warp}
\resizebox{0.99\columnwidth}{!}{
\begin{tabular}{|c|ccc|ccc|}
\hline
Speed Ratio(\%)                                                   & \multicolumn{3}{c|}{Block-Level Partition}                     & \multicolumn{3}{c|}{Combined Warp}                              \\ \hline
\begin{tabular}[c]{@{}c@{}}Column Dimension \\ Range\end{tabular} & \multicolumn{1}{c|}{Avg}   & \multicolumn{1}{c|}{Max}   & Min  & \multicolumn{1}{c|}{Avg}   & \multicolumn{1}{c|}{Max}   & Min   \\ \hline
{[}16, 32{]}                                                      & \multicolumn{1}{c|}{105.2} & \multicolumn{1}{c|}{129.2} & 92.4 & \multicolumn{1}{c|}{133.4} & \multicolumn{1}{c|}{194.5} & 104.8 \\ \hline
(32, 64{]}                                                        & \multicolumn{1}{c|}{107.2} & \multicolumn{1}{c|}{130.7} & 94.1 & \multicolumn{1}{c|}{127.8} & \multicolumn{1}{c|}{174.0} & 87.3  \\ \hline
(64, 96{]}                                                        & \multicolumn{1}{c|}{106.5} & \multicolumn{1}{c|}{127.7} & 92.4 & \multicolumn{1}{c|}{105.5} & \multicolumn{1}{c|}{126.5} & 81.3  \\ \hline
(96, 128{]}                                                       & \multicolumn{1}{c|}{106.8} & \multicolumn{1}{c|}{126.0} & 92.9 & \multicolumn{1}{c|}{122.9} & \multicolumn{1}{c|}{156.0} & 86.7  \\ \hline
\end{tabular}}
\label{tab:speed_ratio}
\end{table}

% \begin{figure}[htbp]
%     \centering
%     \includegraphics[
%     clip, 
%     width =0.9 \linewidth]{figs/two_contribution.pdf} 
%      \caption{
%      Speed ratios of employing block-level partitioning (upper) and combined warp strategies (lower) against not employing them, across all graphs and different column dimensions of the right-hand matrix. 
%      }
%     \label{fig:two_contribution} 
% \end{figure}

\begin{figure}[htbp]
    \centering
    \includegraphics[
    clip, 
    width =0.9 \linewidth]{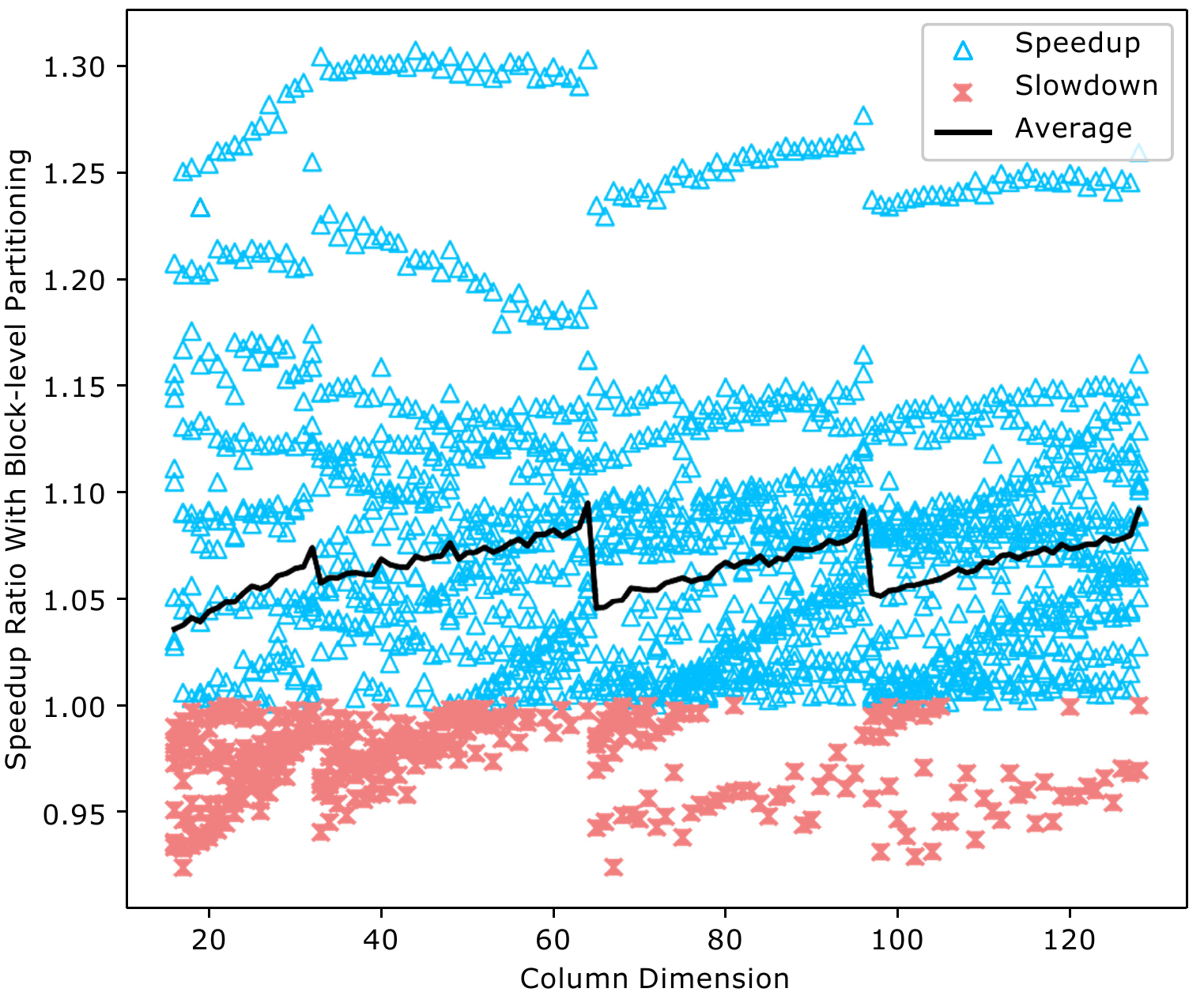} 
     \caption{
     Speedup of (i). degree sorting \& block-level partition over (ii). warp-level partition. Both integrated with combined-warp strategy. 
     }
    \label{fig:contribution_degree_sorting_dynamic_nz_size} 
\end{figure}

%     Speed ratios of employing degree-sorting with dynamic NZ size determining against not employing degree-sorting with fixed NZ size, across all graphs and different column dimensions of the right-hand matrix.

\begin{figure}[htbp]
    \centering
    \includegraphics[
    clip, 
    width =0.9 \linewidth]{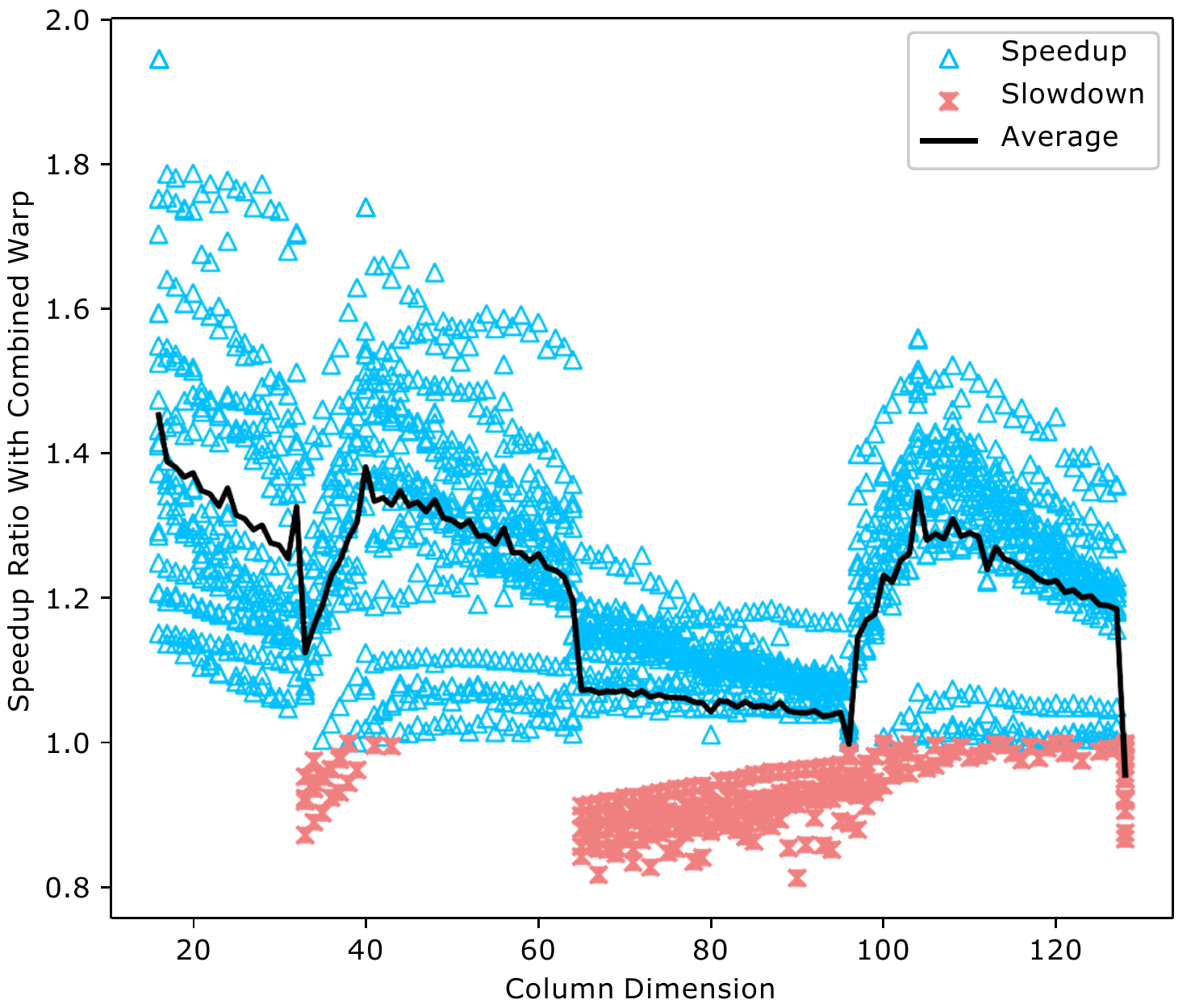} 
     \caption{
     Speedup of degree sorting \& block-level partition (i). with combined-warp strategy over (ii). without combined-warp strategy. 
     }
    \label{fig:contribution_combined_warp} 
\end{figure}

%     Speed ratios of employing combined warp strategies against not employing it, across all graphs and different column dimensions of the right-hand matrix. 

\subsection{Performance Evaluation}

As illustrated in Fig. \ref{fig:normalized_speed}, our proposed kernel depicts a comprehensive enhancement in performance compared to cuSPARSE in the majority of the cases, and distinctly outperforms the state-of-the-art predecessors, namely GNNAdvisor and graph-BLAST, in all instances. When considering the average computational performance of kernels across all column dimensions (ranging from 16 to 128) on a variety of graphs, our kernel manifests an average improvement of $1.17\times$ over cuSPARSE, reaching a maximum increment of $1.45\times$. It surpasses a $1.3\times$ improvement in 22\% of the cases, underperforming compared to cuSPARSE in a single instance and approximately equalling cuSPARSE's performance in another. Our kernel significantly supersedes GNNAdvisor and graph-BLAST across all benchmark graphs, yielding an average speedup of $1.86\times$ and $2.94\times$, and a maximum speedup of $3.41\times$ and $5.02\times$, respectively.

Fig. \ref{fig:runtime} exhibits the runtime of all kernels on each graph for every column dimension of the right-hand matrix. Benefitting from memory coalescing and automatic alignment of intermediate results proffered by the combined warp strategy, the runtime demonstrates a gradual increase as the column dimension escalates, exhibiting minimal effect when the column dimension deviates from a power of 2.

\textbf{Ablation Study 1: Block-level Partition vs. Warp-level Partition}.
Figure~\ref{fig:contribution_degree_sorting_dynamic_nz_size} illustrates the comparative speedup ratio achieved by block-level partition as opposed to warp-level partition. The block-level partition, leveraging dynamically varying NZ group sizes according to node degree, manifests superior shared memory reuse efficiency and enhanced locality relative to warp-level partition. As substantiated by Table~\ref{tab:speed_ratio}, block-level partition has realized an average speedup ranging from $1.05\times$ to $1.07\times$ across disparate column dimension intervals, culminating in a peak improvement of $1.31\times$ and a least effective case of $0.92\times$. Importantly, the enhancement in performance facilitated by block-level partitioning is observed to remain consistent across varying column dimensions.

\textbf{Ablation Study 2: Combined Warp}.
Depicted in Fig.~\ref{fig:contribution_combined_warp}, the speedup resulting from block-level partition (i) with and (ii) without combined warp strategies are given. The implementation of combined warp strategy leads to performance improvement specifically within the column dimension intervals [0, 32], [32, 64], and [96, 128], with an average speed gain recorded between $1.23\times$ and $1.33\times$. Conversely, this enhancement is somewhat diminished within the column dimension range [64, 96], a divergence potentially ascribable to unaligned cache line size in the prevailing GPU architecture.

In summation, the ablation studies collectively attest to the vital contributions of both block-level partition and combined warp strategy in accelerating processing speed for most of the graphs. The nuanced differences between these strategies highlight the necessity of targeted optimization based on specific column dimension intervals and underscore the potential for further investigation and refinement in future work.

\section{Conclusion}

Existing deep learning acceleration design focus on leveraging sparsity in training and inference phase~\cite{peng2022towards, wu2020intermittent, sheng2022larger, kan2021zero, luo2022codg, li2021generic, zhang2022toward, huang2021sparse, li2022makes, kan2022brain, peng2021accelerating, zhang2022algorithm, bao2019efficient, peng2023PASNet, wang2023digital, huang2022automatic, wang2021lightweight, huang2021hmc, kan2022fbnetgen, huang2022dynamic, xiao2019autoprune, peng2021binary, qi2021accelerating, qi2021accommodating, zhang2023accelerating, bao2020fast, peng2022length}. Most of them focuses on FLOPs reduction in algorithm perspective and lacks system-level solution to provide effective speedup. 

In this work, we presents \ourframework, a GPU accelerator architecture addressing workload imbalance and memory access irregularity in GCNs. Incorporating a lightweight, $\mathcal{O}(n)$ preprocessing stage with degree sorting and block-level partition, it optimizes memory utilization and workload distribution. The kernel design further leverages a combined warp strategy for dense column dimension processing, enhancing performance and memory efficiency. \ourframework further improves memory coalescing and alignment to achieve a better memory bandwidth utilization. Evaluated on 18 benchmark graphs, \ourframework outperforms cuSPARSE, GNNAdvisor, and graph-BLAST by $1.17\times$, $1.86\times$, and $2.94\times$ respectively, highlighting its potential in general GCN acceleration applications.

\section{Acknowledgement}
This work was partially funded by the Semiconductor Research Corporation (SRC) Artificial Intelligence Hardware program,  and the UIUC HACC program.

\bibliographystyle{unsrt}
\bibliography{ref}

\end{document}